\documentclass[prd,onecolumn]{revtex4}
\usepackage{dcolumn}
\usepackage{multirow}
\usepackage{graphicx}
\usepackage{amssymb}
\usepackage{bm}
\usepackage{hyperref}
\usepackage{epstopdf}
\usepackage{color}
\usepackage{mathrsfs}
\usepackage{amsmath,amssymb,amsthm}
\usepackage{rotating}
\usepackage{sverb, longtable}
\usepackage{subfigure}
\begin{document}
\title{Can $f(T)$ gravity resolve the $H_0$ tension? }
\author{Deng Wang}
\email{cstar@sjtu.edu.cn}
\affiliation{National Astronomical Observatories, Chinese Academy of Sciences, Beijing, 100012, China}

\author{David Mota}
\affiliation{Institute of Theoretical Astrophysics, University of Oslo, P.O. Box 1029 Blindern, N-0315
	Oslo, Norway}
\begin{abstract}
Motivated by the discrepancy in measurements of $H_0$ between local and global probes, we investigate whether teleparallel gravities could be a better model to describe the present days observations or at least to alleviate the $H_0$ tension. Specifically, in this work we study and place constraints on three popular $f(T)$ models in light of the Planck-2018 CMB data release. We find that the $f(T)$ power-law model can alleviate the $H_0$ tension from $4.4\sigma$ to $1.9\sigma$ level, while the $f(T)$ model of two exponential fail to resolve this inconsistency. Moreover, for the first time, we obtain constraints on the effective number of relativistic species $N_{eff}$ and on the sum of the neutrino masses $\Sigma m_\nu$ in $f(T)$ gravity. We find that the constraints obtained are looser than in $\Lambda$CDM. However, the introduction of massive neutrinos into the cosmological model alleviate the $H_0$ tension for the power-law model. Finally, we find that whether a viable $f(T)$ theory can mitigate the $H_0$ tension depends on the mathematical structure of the distortion factor $y(z,\,b)$. These results could provide a clue for theoreticians to write a more physical-motivated expression of $f(T)$ function.

\end{abstract}
\maketitle

\section{Introduction}
With more extensive surveys at different scales and improved measuring techniques, measurements of late-time cosmic acceleration and growth of gravitational structure have sharpened considerably in recent years \cite{1}. Independent observations from Planck-2018 cosmic microwave background (CMB) radiation have been tighter than before \cite{2,3,4}. Type Ia supernovae (SNe Ia) \cite{5,6} and baryon acoustic oscillations (BAO) \cite{7,8} have been measured up to redshift $z<3$, and we now have obtained data better than $1\%$ precision for $z<1$. Based on several large weak lensing experiments including Kilo-Degree Survey (KiDS) \cite{9}, the Dark Energy Survey (DES) \cite{10}, and the Subaru Hyper-Suprime Camera (HSC) \cite{11}, measurements of effects of dark matter clustering have approached 2$\sim$3$\%$ precision. On one hand, all the above probes verify the correctness of the standard cosmological paradigm, $\Lambda$-cold dark matter ($\Lambda$CDM) model under the framework of general relativity (GR), in describing the evolution of the universe at both small and large scales. On the other hand, the $\Lambda$CDM scenario faces at least two intractable problems, namely the coincidence and fine-tuning problems (see \cite{12}  for details), and at least two tensions emerged from cosmological observations, namely the Hubble constant ($H_0$) and matter fluctuation amplitude ($\sigma_8$) tensions.
The $H_0$ tension is that the indirectly derived Hubble expansion rate from Planck-2018 CMB data release \cite{2} is 4.4$\sigma$ lower than the direct measurement from Hubble Space Telescope (HST) \cite{13}, while the $\sigma_8$ one indicates that the amplitude of density fluctuations today in linear regime, from Planck-2018 data is, nonetheless, higher than the same quantity measured by several low redshift probes including weak gravitational lensing \cite{14}, cluster counts \cite{15} and redshift space distortions \cite{16}. So far, it is still unclear that these tensions are originated from unknown systematic errors in data processing, or new physics beyond $\Lambda$CDM at all? 
Since the $H_0$ tension recently becomes more severe than before \cite{13}, much more attention in the community is paid to alleviating or even solving this large discrepancy. From a point of view of pure theory, except finding out possible systematic uncertainties or using other independent probes to give a resolved determination of $H_0$, we argue that the most direct way is to check the model dependence of Planck-2018 CMB data. Along this line, a great deal of effort has been implemented by cosmologists under the hypothesis of dark energy or equivalently modified gravity \cite{17,18,19,20,21,22,23,24,25,26,27}.           

In this work, we are motivated by exploring that whether the teleparallel equivalent of GR \cite{28} can resolve current $H_0$ tension. Starting from the Lagrangian, the simplest representative of teleparallel gravity is $f(T)$ gravity \cite{29}, which is completely equivalent to GR at the level of equations. Since $f(T)$ gravity is firstly proposed \cite{30}, many authors have placed constraints on its extensions using the cosmological observations \cite{31,32,33,34,a1,a2,a3}. However, the question is that CMB data is always combined with BAO, SNe Ia, local $H_0$ observation and other probes to implement strict constraints. More or less, this kind of constraint can only provide the indirect test of $H_0$ tension in the framework of $f(T)$ gravity. Therefore, there is still a lack of a direct test of the ability to resolve the $H_0$ tension for $f(T)$ gravity in light of Planck CMB data. Especially, after the final data release of Planck-2018 full mission, this is an urgent issue needed to be addressed. By implementing numerical analysis, we find that the power-law $f(T)$ gravity can efficiently resolve current $H_0$ tension, but the exponential $f(T)$ gravity fails to do this.         

This work is outlined in the following manner. In the next section, we introduce the formalism of $f(T)$ gravity and specify three $f(T)$ gravity models to be constrained by cosmological observations. In Section III, we describe the data and methodology used in this analysis. In Section IV, we display our numerical results and discussions. The conclusions are presented in the final section. 

\section{$f(T)$ cosmological models}
The dynamical variable of $f(T)$ gravity is the vierbein field $\mathbf{e}_A^\mu$, which constructs an orthonormal basis for the tangent space at each point $x^\mu$ of the space-time manifold $M$. Note that here we, respectively, use Greek and capital Latin indices to denote the space-time coordinates and the coordinates of the tangent space. Utilizing the components of vierbein vector, the metric in $f(T)$ gravity can be written as $g_{\mu\nu}=\eta_{AB}e_\mu^Ae_\nu^B$, where $\eta_{AB}$ is the Minkowski metric for the tangent space at each $x^\mu$. Furthermore, through replacing the nonzero-curvature Levi-Civita connection with the torsional Weitzenb\"{o}ck one \cite{35}, one can express the torsion tensor as      
\begin{equation}
T^\gamma_{~~\mu\nu}\equiv e^\gamma_{~A}(\partial_\mu e^A_{~\nu}-\partial_\nu e^A_{~\mu}). \label{1}
\end{equation}
By contractions of the torsion tensor, the torsion scalar $T$ in the Lagrangian density can be shown as 
\begin{equation}
T\equiv\frac{1}{4}T^{\gamma\mu\nu}T_{\gamma\mu\nu}+\frac{1}{2}T^{\gamma\mu\nu}T_{\nu\mu\gamma}-T_{\gamma\mu}^{~~\gamma} T^{\nu\mu}_{~~~\nu}.    \label{2}
\end{equation} 
Very similar to the case of $f(R)$ gravity, the idea of $f(T)$ gravity
is to generalize $T$ to an arbitrary function $f(T)$, when the action is constructed by the teleparallel Lagrangian density $T$. Specifically, the action of $f(T)$ gravity in a universe can be written as 
\begin{equation}
\mathcal{S} = \int d^4x\,|e|\,\frac{T+f(T)}{16\pi G}+\mathcal{S_\mathit{(m)}}, \label{3}
\end{equation} 
where $|e|=\sqrt{-g}$ and $\mathcal{S_\mathit{(m)}}$ denotes the matter field.
One can easily find that GR is recovered when $f(T)=0$ and GR with a cosmological constant is restored when $f(T)=\mathit{const.}$. Varying Eq.(\ref{3}) with respect to the vierbein field $\mathbf{e}_A^\mu$, the field equations of $f(T)$ can be obtained as 
\begin{equation}
e^{-1}\partial_\mu(ee^\gamma_{~A}S_\gamma^{~\mu\nu})(1+f_T)+e^\gamma_{~A}
S_\gamma^{~\mu\nu}\partial_\mu(T)f_{TT}-e^\lambda_{~A}T^\gamma_{\mu\lambda}S_{\gamma}^{~\nu\mu}(1+f_T)+\frac{1}{4}e^\nu_{~A}\left[T+f(T)\right]=4\pi G e^\gamma_{~A}\mathcal{T_\mathit{(m)}}_\gamma^{~\nu}, \label{4}
\end{equation}
where $f_T \equiv \partial f/\partial T$, $f_{TT}\equiv\partial^2f/\partial T^2$, and $\mathcal{T_\mathit{(m)}}_\gamma^{~\nu}$ denote the energy-momentum tensor of matter fields including baryons, dark matter and radiation in the universe.   

If the background space-time manifold is a spatially flat, homogeneous and isotropic one, using the vierbein form $e^A_\mu=\mathrm{diag}(1,~a,~a,~a)$, one shall naturally obtain a Friedmann-Robertson-Walker (FRW) metric
\begin{equation}
ds^2=dt^2-a^2{(t)}\delta_{ij}dx^idx^j,
\end{equation} 
where $t$ and $a$ denote the cosmic time and the scale factor of the universe, respectively. Substituting the chosen vierbein into Eq.(\ref{4}), the Friedmann equations of $f(T)$ gravity reads
\begin{equation}
3H^2=8\pi G(\rho_b+\rho_{cdm}+\rho_r)+Tf_T-\frac{f}{2}, \label{6}
\end{equation}
\begin{equation}
\dot{H}=-\frac{4\pi G(\rho_b+\rho_{cdm}+\rho_r+P_b+P_{cdm}+P_r)}{2Tf_{TT}+f_T+1}, \label{7}
\end{equation}
where $\rho_i$ and $P_i$ $(i=b, ~cdm, ~r)$ denote the energy densities and pressures of different matter components including baryons ($b$), cold dark matter ($cdm$) and radiation ($r$). $H$ is Hubble parameter and the dot represents the derivative with respect to the cosmic time $t$. Different from the case of $f(R)$ gravity, we have a more elegant expression between Hubble parameter $H$ and torsional scalar $T$
\begin{equation}
T=-6H^2, \label{8}
\end{equation}  
which can be naturally derived from Eq.(\ref{2}) in the FRW vierbein. At the present time, this simple relation reads $T_0=-6H^2_0$. As a consequence, we have dimensionless Hubble parameter $E^2(z)\equiv H^2(z)/H^2_0=T/T_0$.

It is not difficult to see that the latter two terms in Eq.(\ref{6}) is responsible for explaining the cosmic acceleration. The torsional fluid can be regarded as an effective dark energy fluid. Hence, one can obtain the effective energy density $\rho_{de}$ and pressure $P_{de}$ of dark energy as     
\begin{equation}
\rho_{de}= \frac{1}{16\pi G}\left(  2Tf_T-f \right), \label{9}
\end{equation} 
\begin{equation}
P_{de}= \frac{1}{16\pi G}\left(   \frac{2T^2f_{TT}-Tf_T+f}{2Tf_{TT}+f_T+1} \right). \label{10}
\end{equation}
As a consequence, the effective equation of state (EoS) of dark energy $\omega_{de}$ is written as    
\begin{equation}
\omega_{de}=   \frac{2T^2f_{TT}-Tf_T+f}{(2Tf_T-f)(2Tf_{TT}+f_T+1)}, \label{11}
\end{equation} 
Subsequently, since matter and dark energy are independent components in the dark sector of the universe, the energy conservation equation for dark energy can also be shown as 
\begin{equation}
\dot{\rho_{de}}+3H(1+\omega_{de})\rho_{de}=0. \label{12}
\end{equation}   

In order to perform constraints on $f(T)$ gravity models using data, one can rewrite Eq.(\ref{6}) in the following manner
\begin{equation}
E^2(z)=\Omega_{m0}(1+z)^3+\Omega_{r0}(1+z)^4+(1-\Omega_{m0}-\Omega_{r0})y(z,\mathbf{w}), \label{13}
\end{equation}
where $\Omega_{m0}$ and $\Omega_{r0}$ are, respectively, the present-day values of matter and radiation densities. The factor $y(z, \mathbf{w})= (T_0-2Tf_T)/[T_0(1-\Omega_{m0}-\Omega_{r0})]$ \cite{32}, where $\mathbf{w}$ is a set of typical parameters of a specific $f(T)$ model, characterizes the modification effect of $f(T)$ gravity relative to $\Lambda$CDM.     
  
An underlying and subtle rule to construct an alternative cosmological model is that this new model can be reduced to $\Lambda$CDM when its typical parameter takes some certain value. For instance, $\omega$CDM model comes back to $\Lambda$CDM when the EoS of perfect dark energy fluid $\omega=-1$. Similarly, we will consider this kind of $f(T)$ models in our treatment.

\begin{figure}
	\centering
	\includegraphics[scale=0.7]{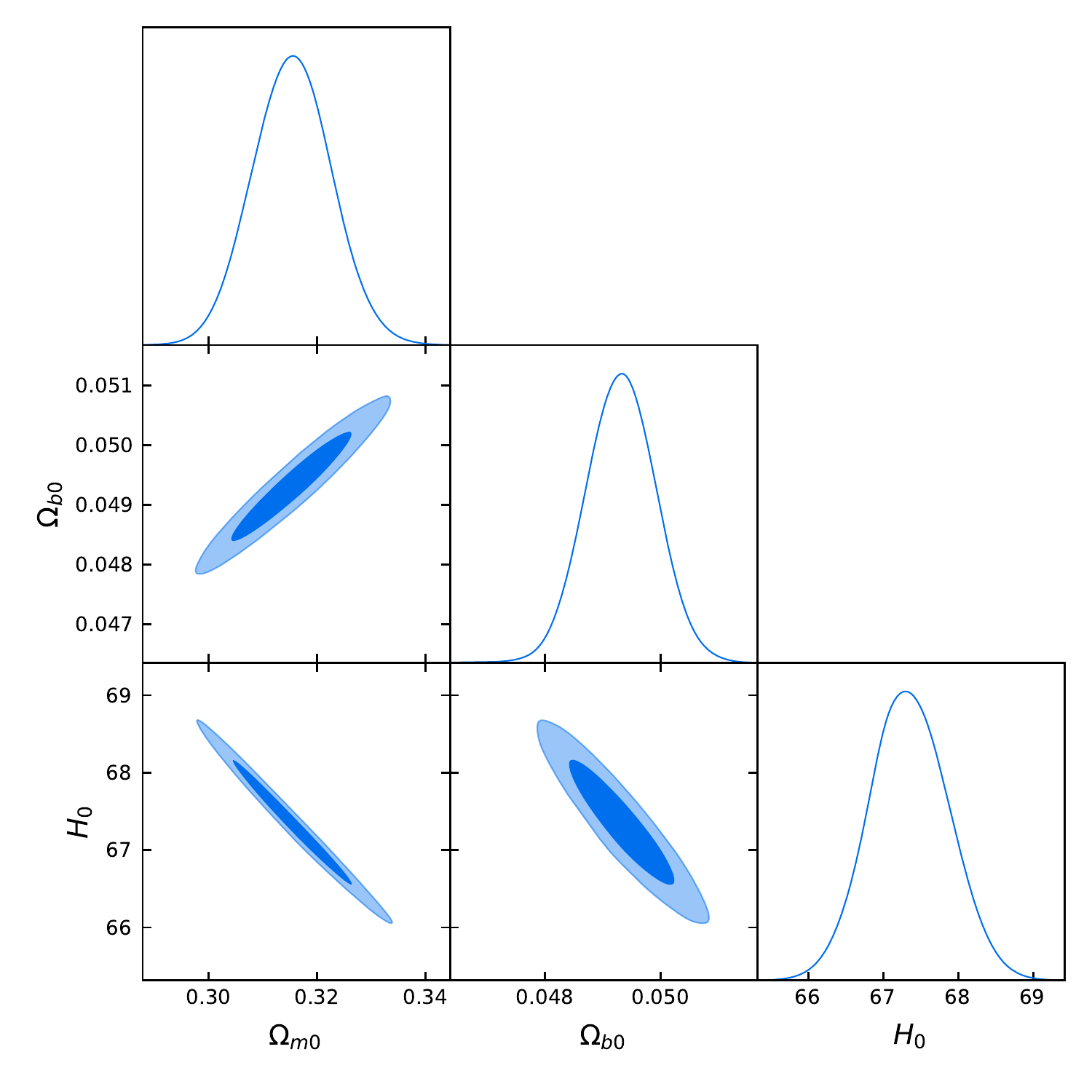}
	\caption{Marginalized $1\sigma$ ($68\%$) and $2\sigma$ ($95\%$) constraints on the $\Lambda$CDM model using the Planck-2018 CMB data. }\label{f1}
\end{figure}
\begin{figure}
	\centering
	\includegraphics[scale=0.7]{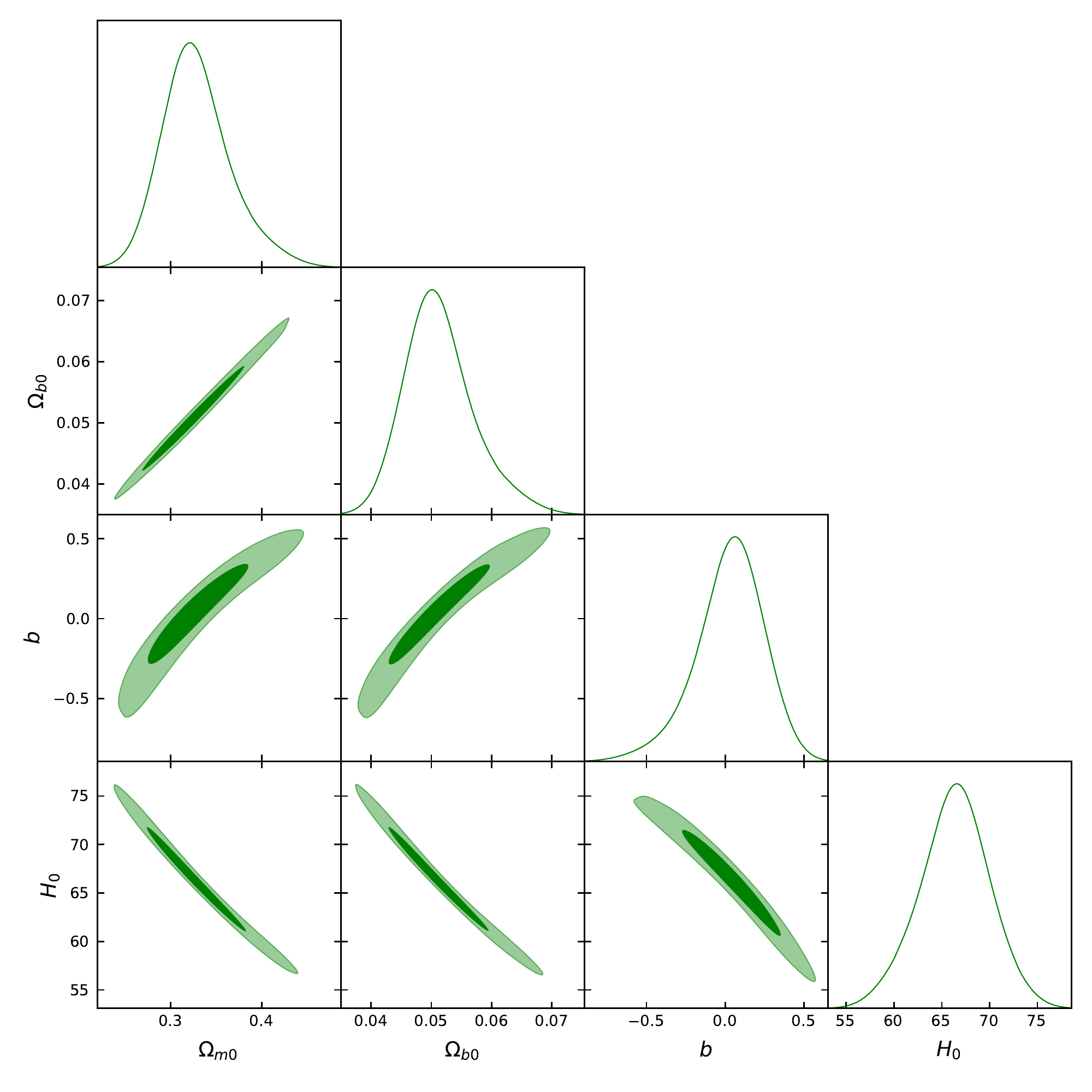}
	\caption{Marginalized $1\sigma$ ($68\%$) and $2\sigma$ ($95\%$) constraints on the M1 model using the Planck-2018 CMB data. }\label{f2}
\end{figure}
\begin{figure}
	\centering
	\includegraphics[scale=0.7]{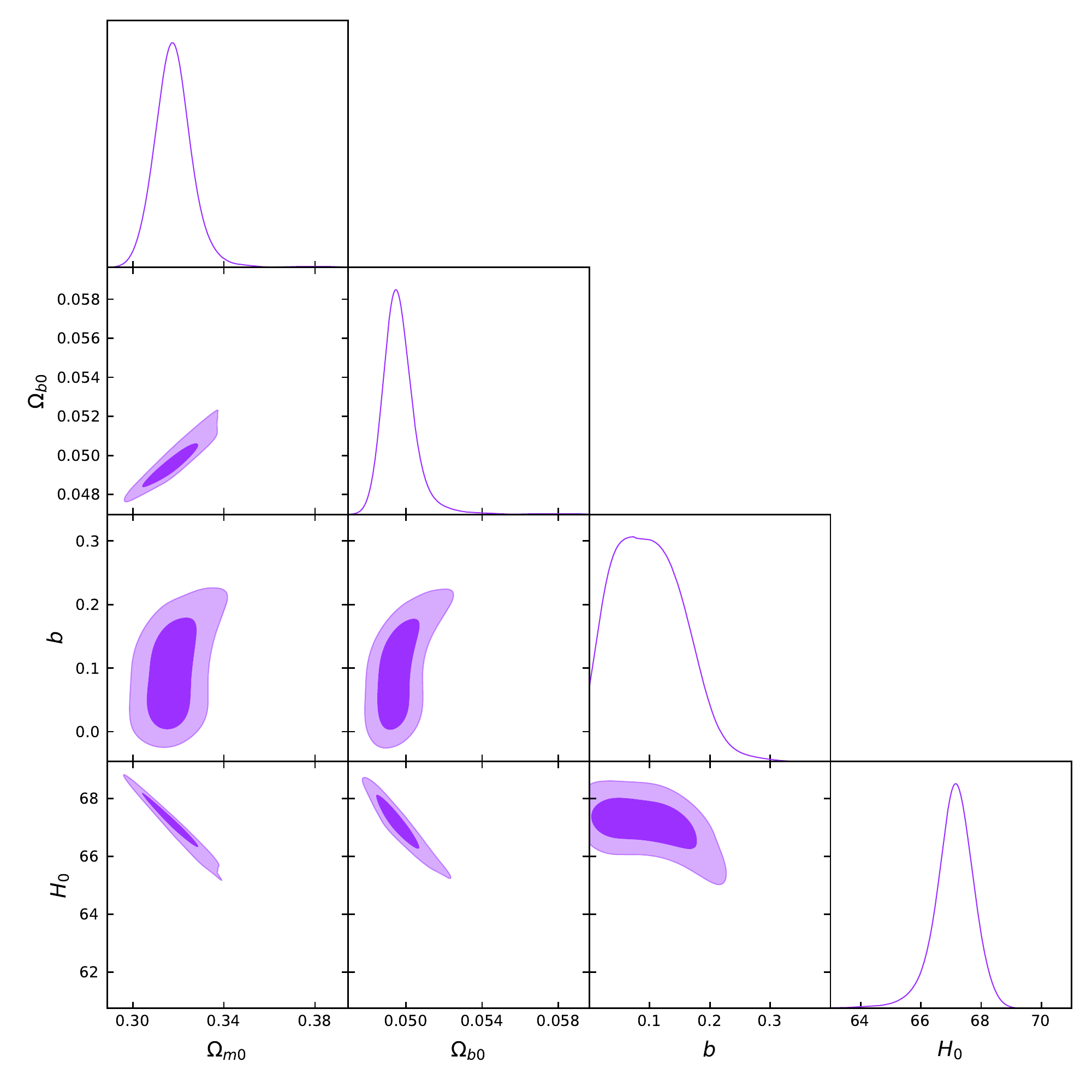}
	\caption{Marginalized $1\sigma$ ($68\%$) and $2\sigma$ ($95\%$) constraints on the M2 model using the Planck-2018 CMB data. }\label{f3}
\end{figure}
\begin{figure}
	\centering
	\includegraphics[scale=0.7]{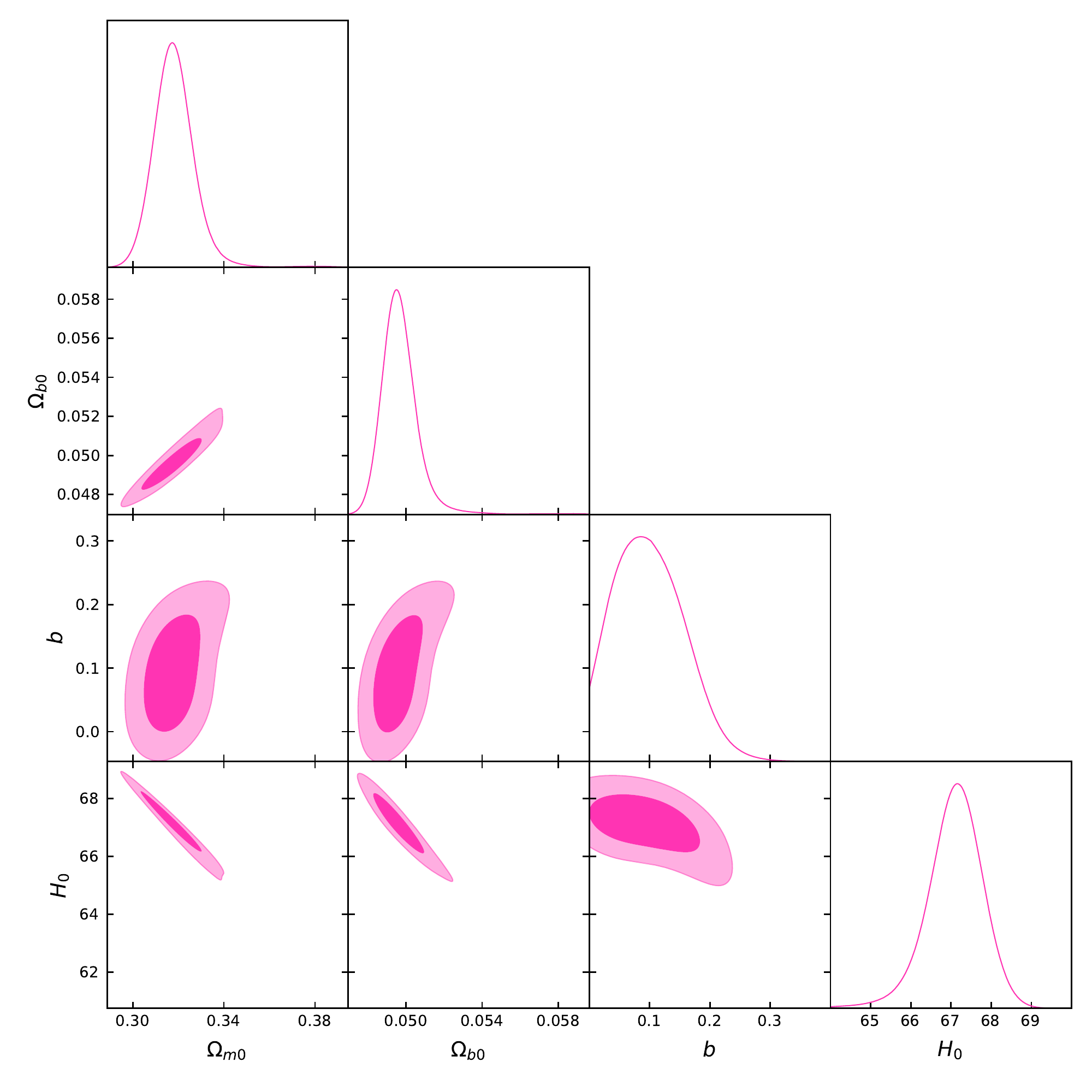}
	\caption{Marginalized $1\sigma$ ($68\%$) and $2\sigma$ ($95\%$) constraints on the M3 model using the Planck-2018 CMB data. }\label{f4}
\end{figure}
\begin{figure}
	\centering
	\includegraphics[scale=0.5]{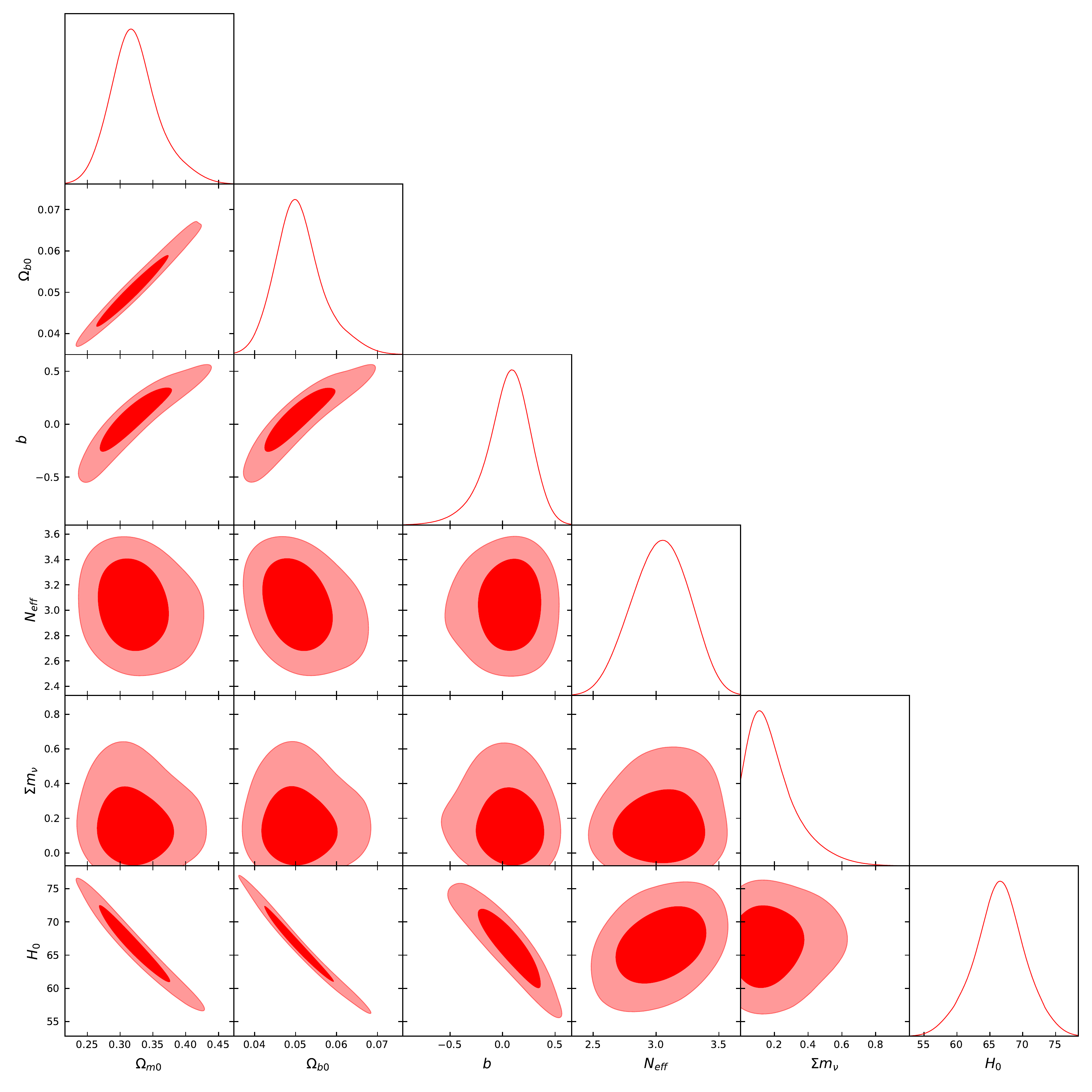}
	\caption{Marginalized $1\sigma$ ($68\%$) and $2\sigma$ ($95\%$) constraints on the M1$\nu$ model using the Planck-2018 CMB data. }\label{f5}
\end{figure}
\begin{figure}
	\centering
	\includegraphics[scale=0.5]{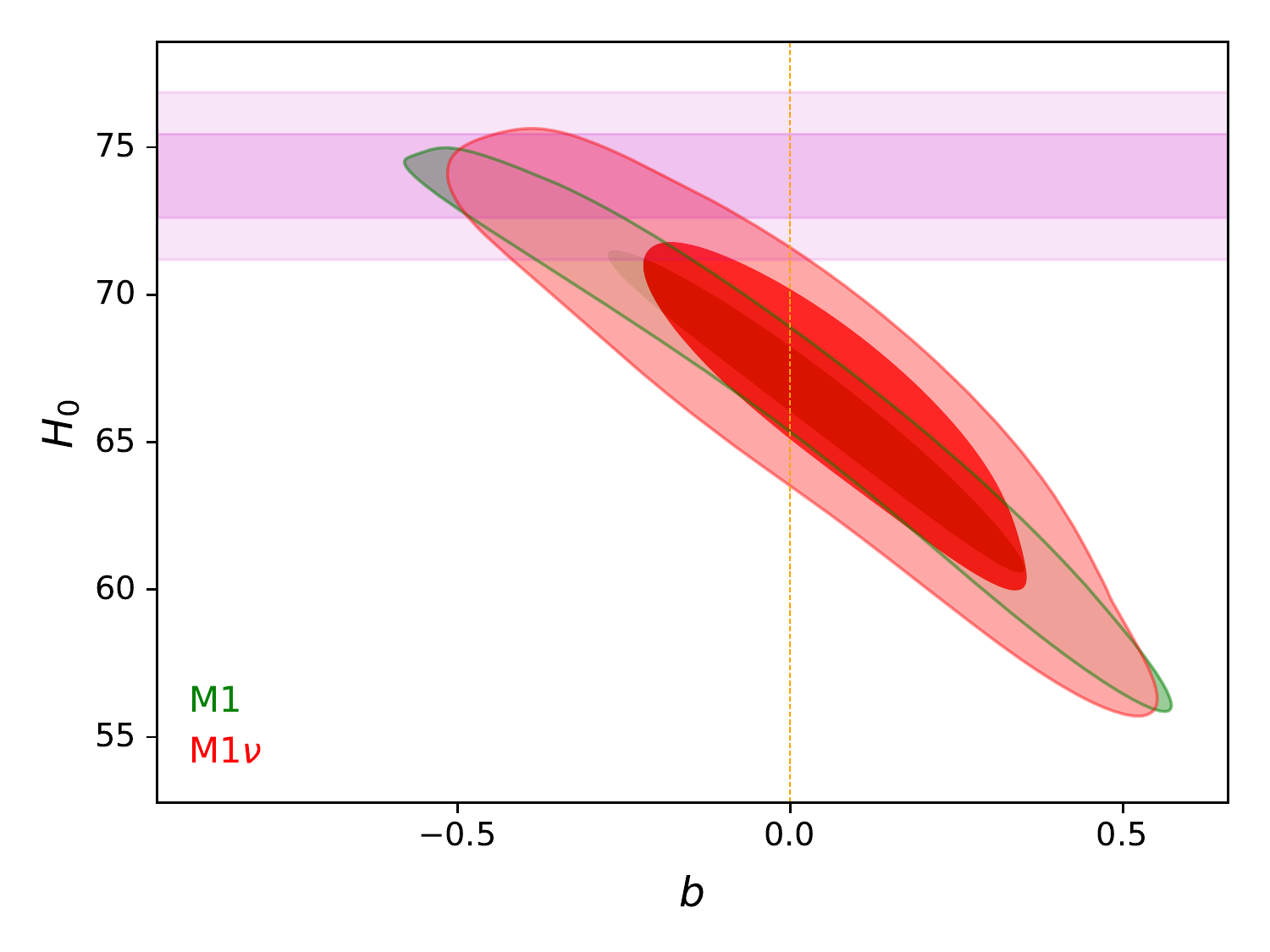}
	\caption{The $H_0$-$b$ plane in the M1 (green) and M1$\nu$ (red) models constrained by the Planck-2018 CMB data. The magenta bands represents the direct measurement $H_0=74.03\pm1.42$ km s$^{-1}$ Mpc$^{-1}$ from the HST project \cite{13}, while the orange line is $b=0$ corresponding to the $\Lambda$CDM case.  }\label{f6}
\end{figure}

In order to investigate whether $f(T)$ gravity can alleviate the $H_0$ tension, specifically, we will constrain three $f(T)$ alternatives commonly used in the literature, which can successfully pass the constraints from the solar system and produce the late-time cosmic acceleration well. These models are still alive in light of current cosmological observations. For the convenience of expression, we use a universal parameter $b$ to rewrite the modification factor as $y(z,b)$. 

$\bullet$ In order to obtain an accelerated expansion without invoking dark energy but driven by torsion, the authors in Ref.\cite{30} proposed a simple power-law model (hereafter M1)
\begin{equation}
f(T)=\alpha(-T)^b, \label{14}
\end{equation}
where $\alpha$ and $b$ denote two free parameters, but only one is independent. Substituting the above expression into Eq.(\ref{6}), one can easily obtain 
\begin{equation}
\alpha=\frac{(1-\Omega_{m0}-\Omega_{r0})(6H_0^2)^{1-b}}{2b-1}, \label{15}
\end{equation} 
and get the corresponding factor 
\begin{equation}
y(z,b)=E^{2b}(z,b). \label{16}
\end{equation} 
It is noteworthy that, for this model, the necessary limitation $b<1$ corresponds to the cosmic acceleration, and that the $\Lambda$CDM scenario recovers when $b=0$. 

$\bullet$ In order to keep the variation of the gravitational coupling small within $f(T)$ theory, Linder also proposed an exponential model (hereafter M2) by analogy with his exponential $f(R)$ gravity \cite{36}, which is shown as    
\begin{equation}
f(T)=\xi\, T_0(1-e^{-p\sqrt{T/T_0}}), \label{17}
\end{equation}
where $xi$ and $p$ are two parameters. In the same light, $c$ can be expressed as 
\begin{equation}
\xi=\frac{1-\Omega_{m0}-\Omega_{r0}}{1-(1+p)e^{-p}}, \label{18}
\end{equation}
and consequently, after some algebraic manipulations,  the modification factor is written as 
\begin{equation}
y(z,b)=\frac{1-(1+\frac{E}{b})e^{-\frac{E}{b}}}{1-(1+\frac{1}{b})e^{-\frac{1}{b}}}, \label{19}
\end{equation}
where $p=1/b$. It is easy to see that M2 is reduced to $\Lambda$CDM when the distortion parameter $b\rightarrow0^{+}$ and GR is recovered when $b\rightarrow+\infty$.  

$\bullet$ Similar to M2 inspired by exponential $f(R)$ gravity, Bamba {\it et al.} \cite{37} also proposed another exponential model (hereafter M3)
\begin{equation}
f(T)=\eta\, T_0(1-e^{-qT/T_0}), \label{20}
\end{equation}
where $\eta$ and $q$ denote two parameters. Similarly, one can have
\begin{equation}
\eta=\frac{1-\Omega_{m0}-\Omega_{r0}}{1-(1+2q)e^{-q}}, \label{21}
\end{equation} 
\begin{equation}
y(z,b)=\frac{1-(1+\frac{2E^2}{b})e^{-\frac{E^2}{b}}}{1-(1+\frac{2}{b})e^{-\frac{1}{b}}}, \label{22}
\end{equation}
where $q=1/b$. One can easily find that M2 and M3 has almost same $f(T)$ structures and distortion factors $y(z,b)$. Therefore, M3 also exhibits same behaviors when $b\rightarrow0^{+}$ or $+\infty$.  

It is worth noting that these models we consider can effectively avoid the Lorentz non-invariance problem and pass the solar system test \cite{32}, since they can be reduced to $\Lambda$CDM when the key parameter $b\rightarrow0$.

The cosmological perturbations in the framework of $f(T)$ gravity are first investigated in Ref.\cite{b1}, where the authors derive the gauge-invariant perturbation equations and study the large scale structure for a specific $f(T)$ model. In Ref.\cite{b2}, the authors generalize the effective field theory approach to torsional modified gravity, which is a formalism that allows for the systematic investigation of the background and perturbation levels separately. Most recently, full sets of linear perturbation equations in $f(T)$ gravity are also derived in Ref.\cite{b3}. In this analysis, we would like to focus on the background evolution of the universe in $f(T)$ gravity.

Using the above mentioned rule to construct a viable $f(T)$ model with more parameters may be a good solution to alleviate or even solve the $H_0$ tension. However, an elegant cosmological model should have parameters as few as possible. As a consequence, the most important for us is to check whether these three $f(T)$ models (M1, M2 and M3) with two parameters in hand can resolve such a large $H_0$ discrepancy.   

\begin{figure}
	\centering
	\includegraphics[scale=0.6]{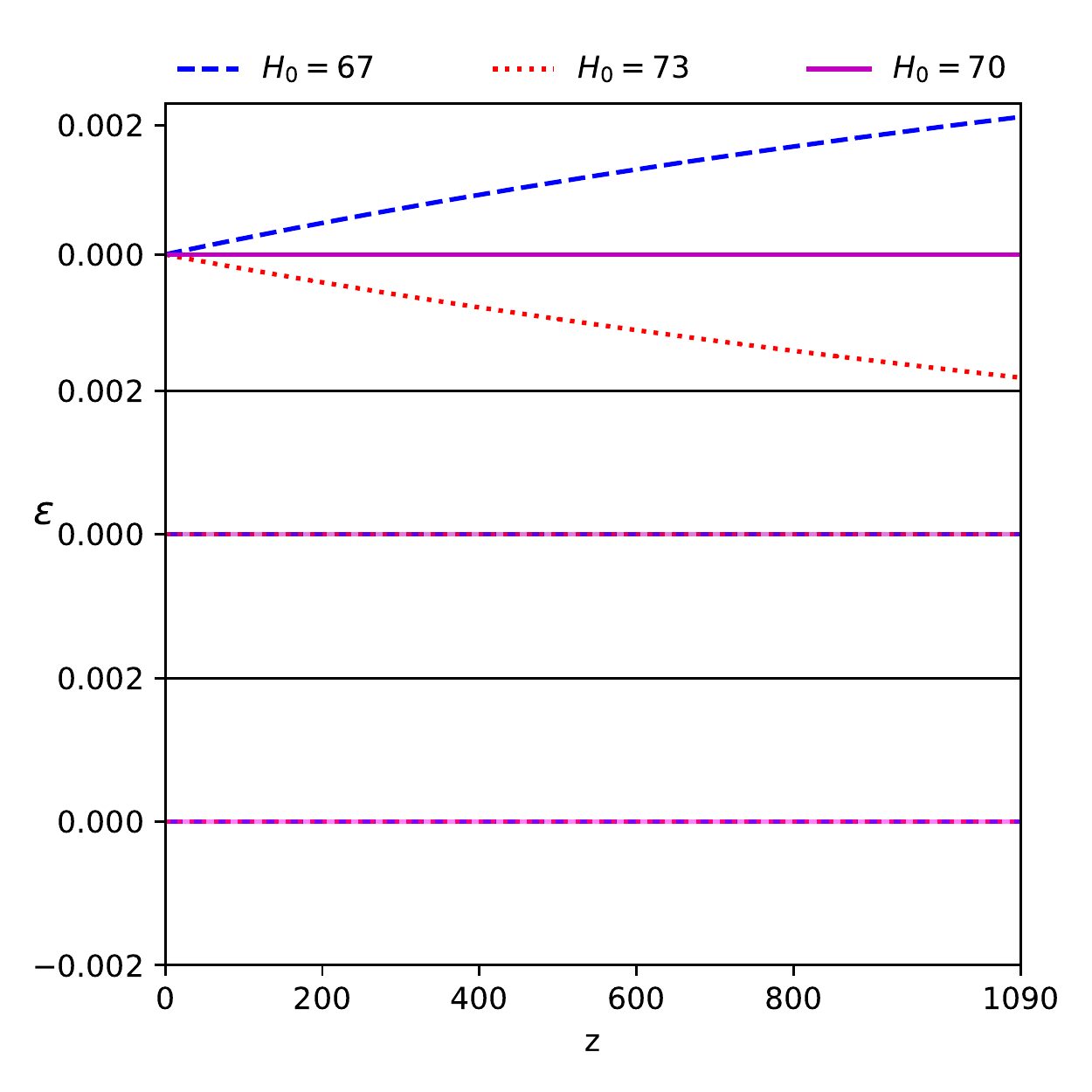}
	\caption{The relation between the relative difference of distortion factor $\epsilon$ and redshift $z$ for M1 (top), M2 (medium) and M3 (bottom). The magenta solid, blue dashed and red dotted lines denote the $\epsilon$-$z$ relations when $H_0=70,\,67$ and 73 km s$^{-1}$ Mpc$^{-1}$, respectively. For three $f(T)$ models, we have assumed $b=0.1$ and $\Omega_{m0}=0.3$.    }\label{f7}
\end{figure}

\begin{figure}
	\centering
	\includegraphics[scale=0.55]{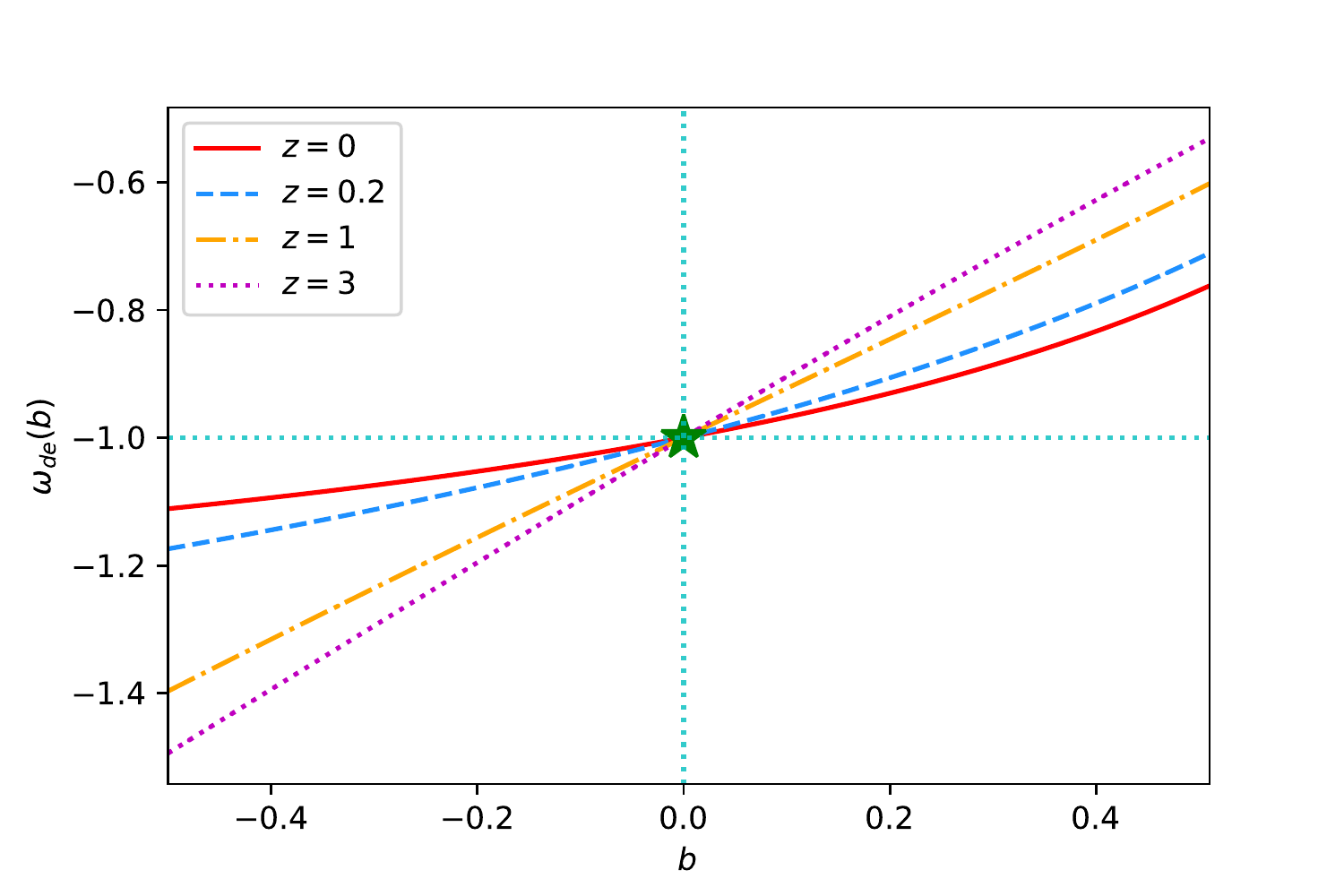}
	\includegraphics[scale=0.55]{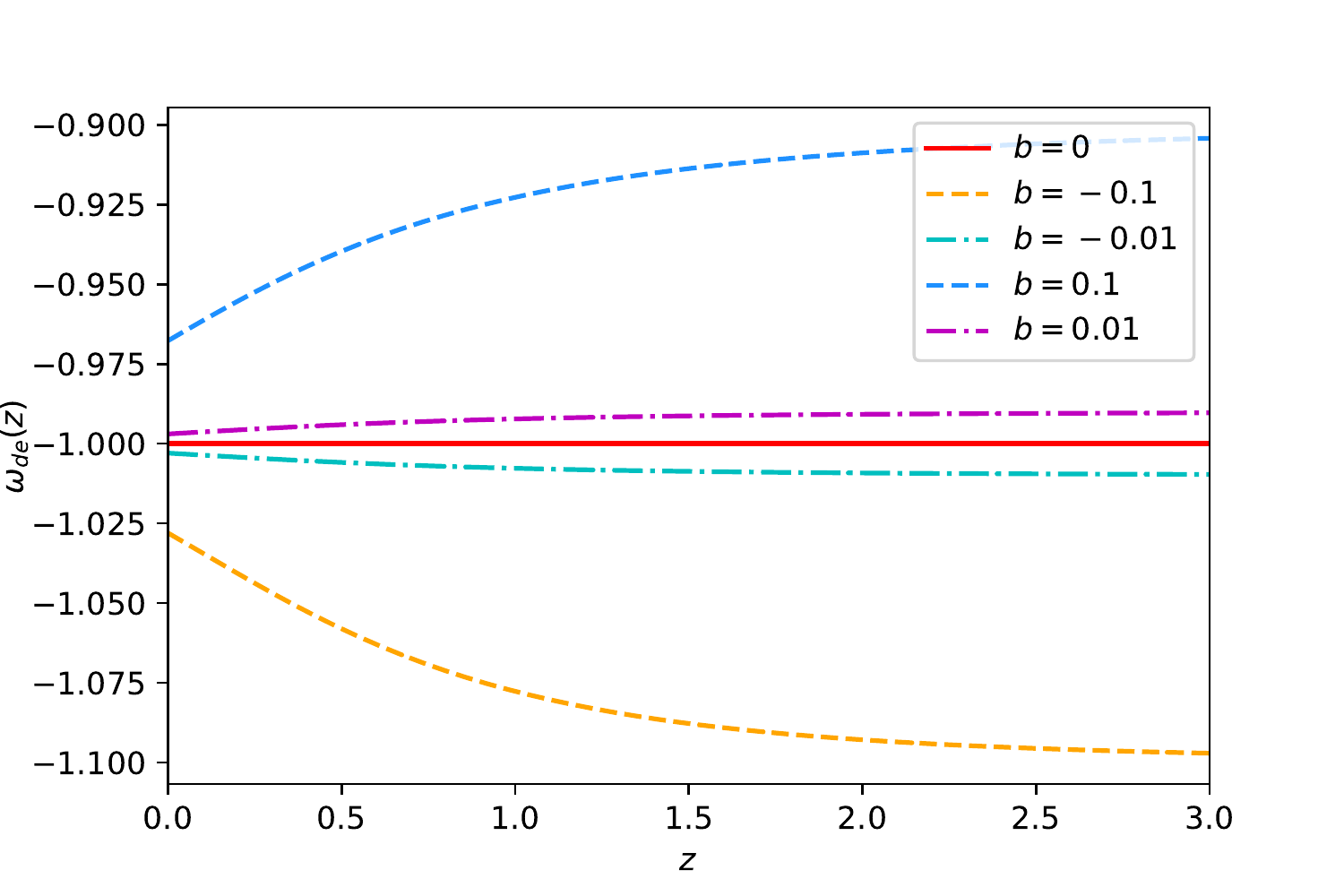}
	\includegraphics[scale=0.55]{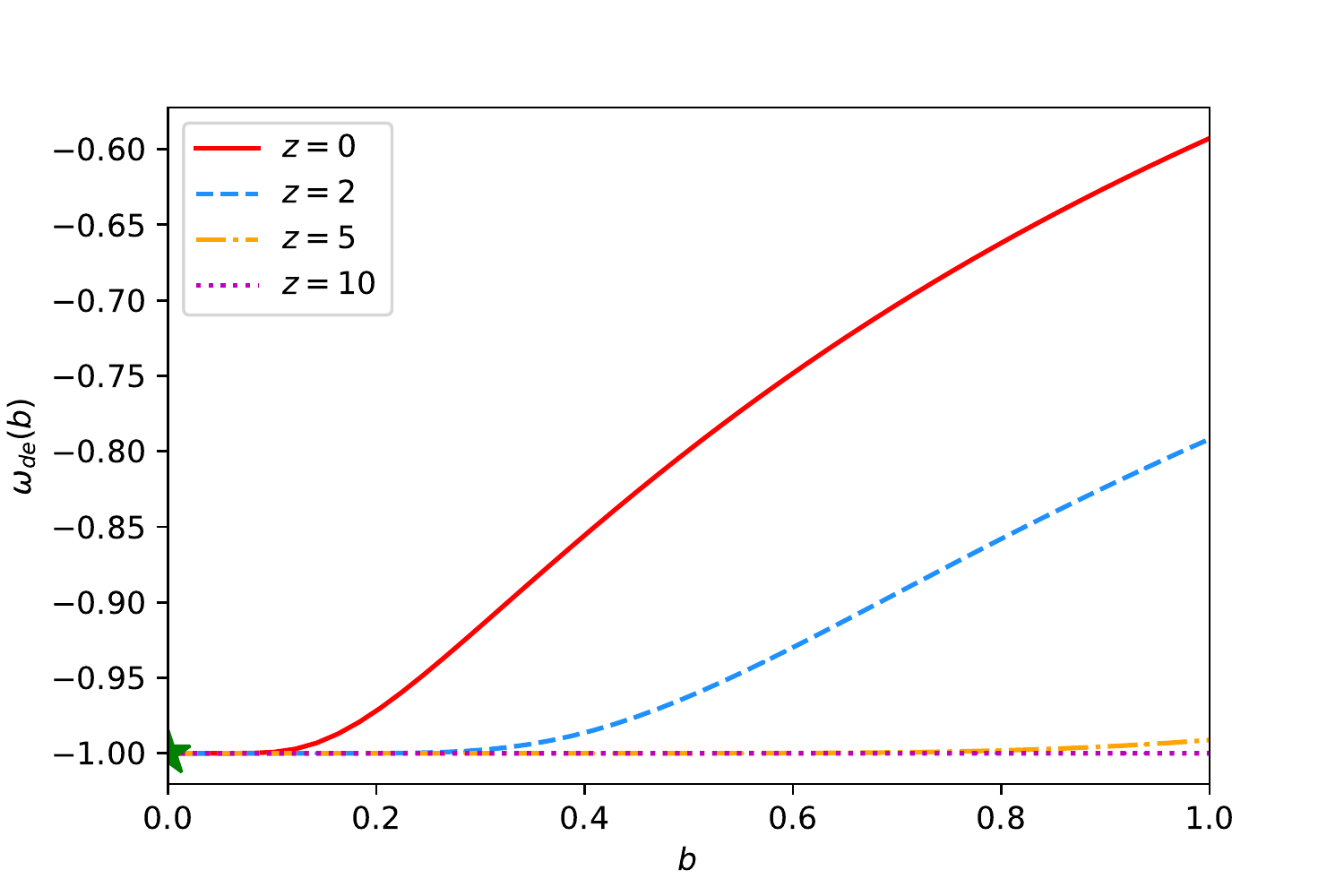}
	\includegraphics[scale=0.55]{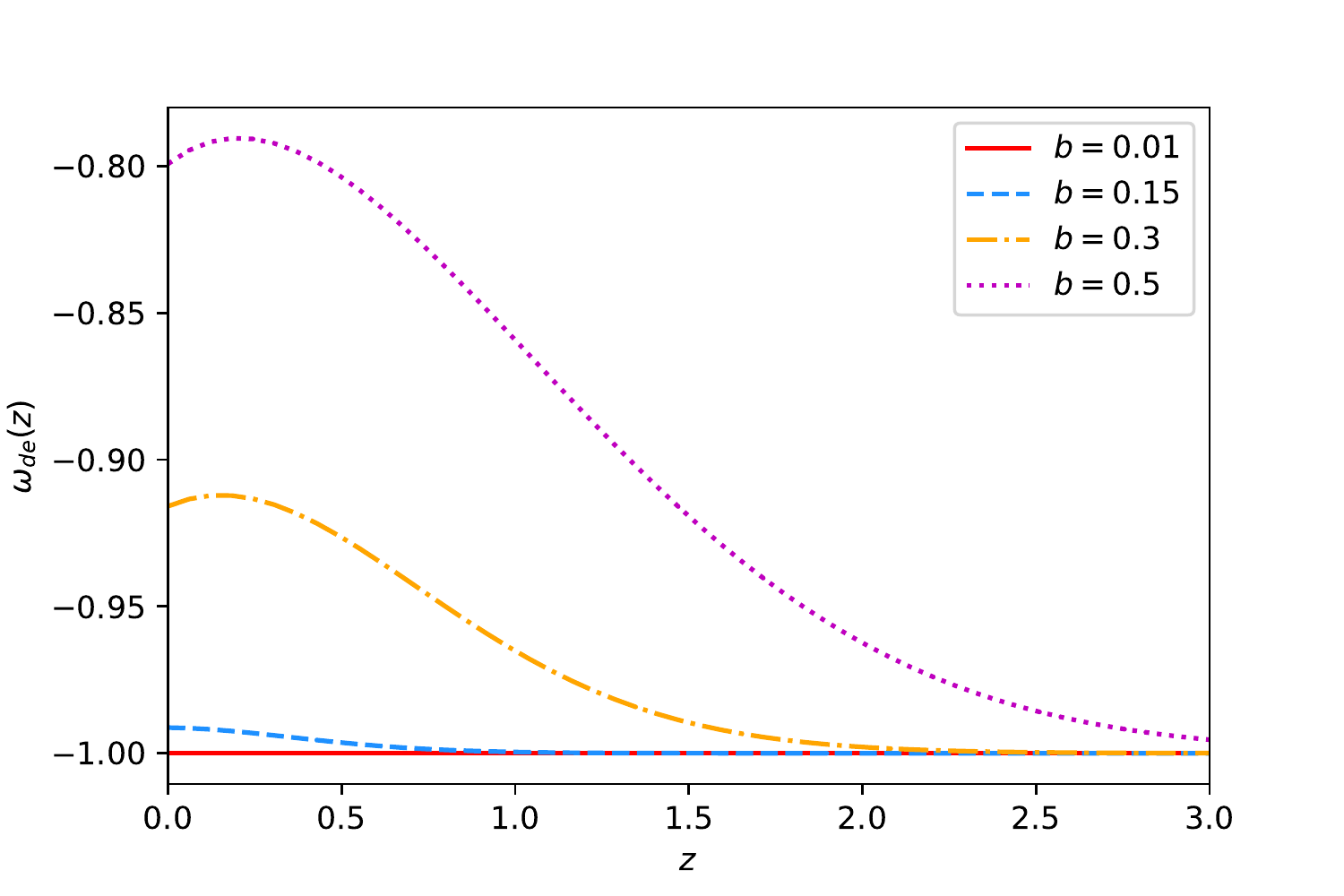}
	\includegraphics[scale=0.55]{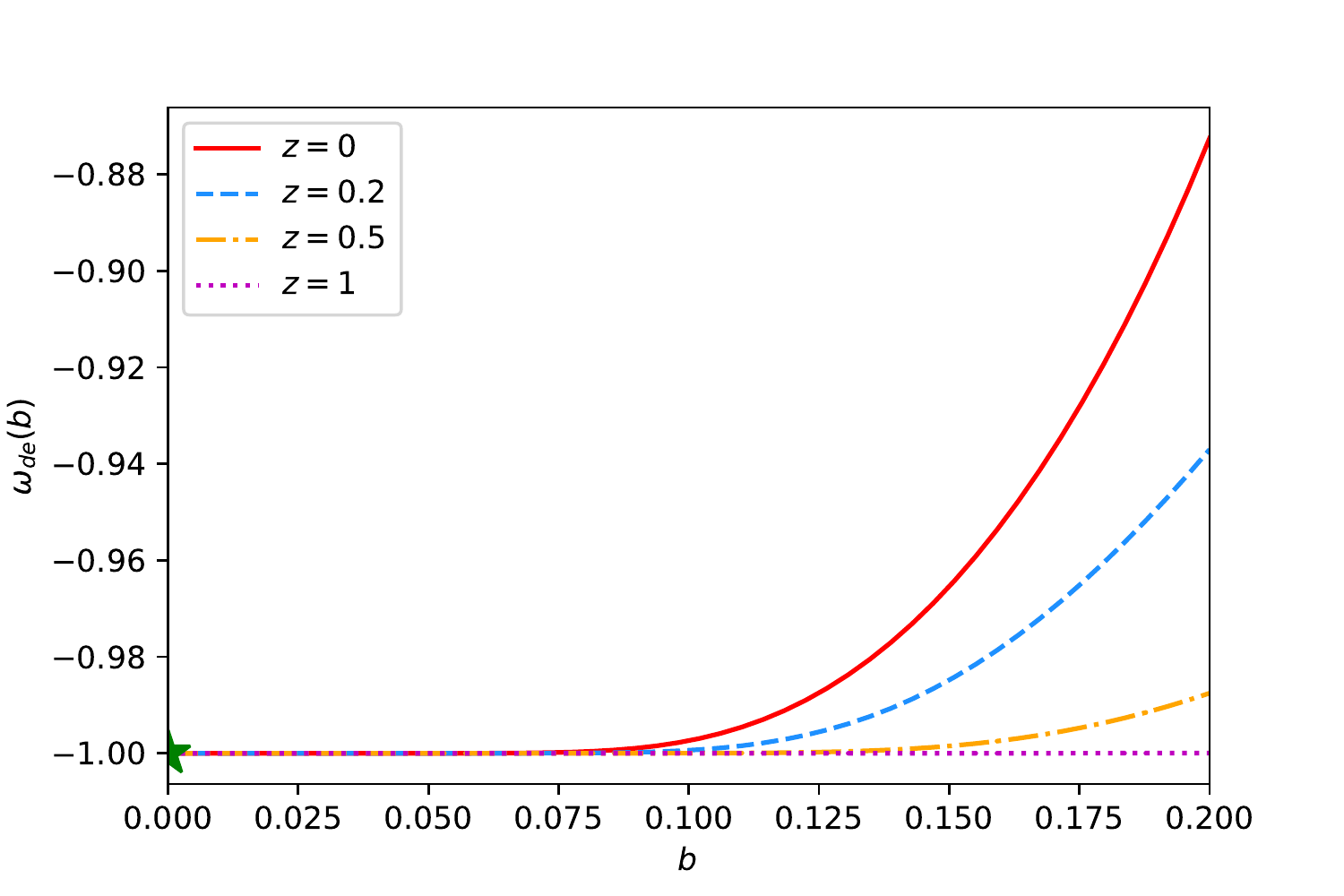}
	\includegraphics[scale=0.55]{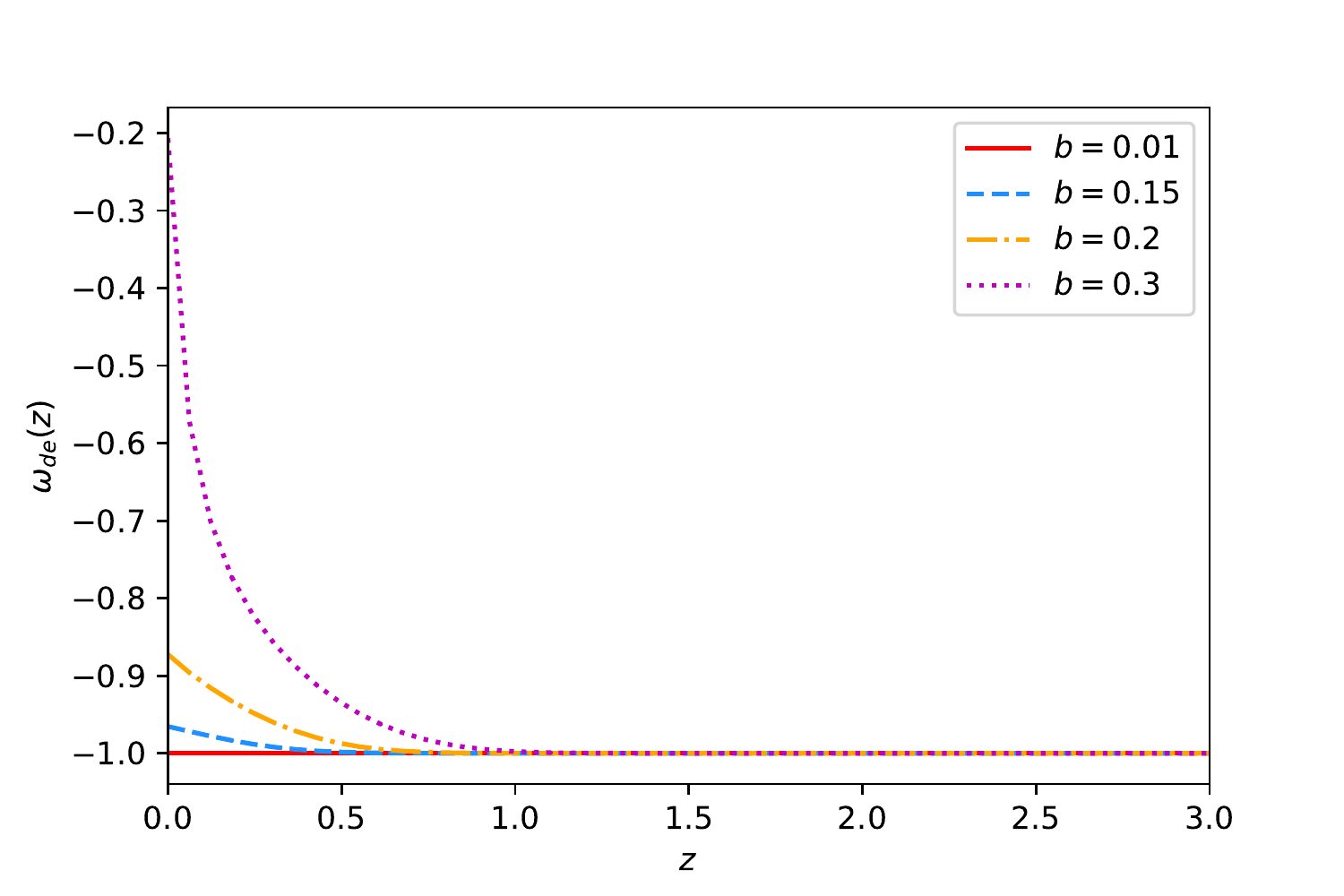}
	\caption{The effective equation of state of dark energy $\omega_{de}$ as a function of the model parameter $b$ or redshift $z$ are shown for M1 (top), M2 (medium) and M3 (bottom) models, respectively. The symbols ``$\star$'' denote the $\Lambda$CDM model in all cases. For three $f(T)$ models, we have assumed $H_0=70$ km s$^{-1}$ Mpc$^{-1}$ and $\Omega_{m0}=0.3$. }\label{f8}
\end{figure}

\section{Data and Method}
As mentioned above, the most straightforward way to test the ability of a model in resolving $H_0$ tension is to investigate its model dependence on CMB data. Hence, we shall constrain these three models by using final Planck-2018 CMB  data release. In principle, one should utilize the original CMB temperature and polarization data to directly constrain these alternatives. Based on the fact that $H_0$ is only sensitive to the distance information extracted from CMB data, one can also use the distance-related information instead. In order to save computational effort and improve the investigation efficiency, in this analysis, we would like to use the distance prior from TTTEEE$+$lowl$+$lowE$+$lensing data, i.e., compressed CMB data obtained in Ref.\cite{38} to implement constraints on $f(T)$ gravity.   

Compared to the Planck-2015 results, in the Planck-2018 release, improved measurements of large-scale polarization and improved modelling of small-scale polarization lead to better constraints on cosmological parameters. The contribution from CMB data in likelihood analysis can be expressed with the corresponding shift parameters 
\begin{equation}
R = \frac{r(z_\star)H_0 \sqrt{\Omega_{m0}}}{c}, \label{23}
\end{equation}
\begin{equation}
l_a = \frac{ r(z_\star)\pi}{r_s(z_\star)}, \label{24}
\end{equation}
where $c$ is the speed of light, $r(z)$ is the comoving distance at redshift $z$, $r_s(z)$ is the comoving sound horizon at $z$, and $z_\star$ is the redshift to the photon-decoupling surface. These two parameters combined with baryon density $\omega_b=\Omega_{b0}h^2$ ($h\equiv H_0/100$ km s$^{-1}$ Mpc$^{-1}$) and the spectral index of primordial power spectrum $n_s$ can provide a brief and efficient extraction from full CMB data for us to implement constraints on dark energy. The comoving sound horizon $r_s(z)$ reads
\begin{equation}
r_s(z) = \frac{c}{H_0}\int_{0}^{a}\frac{d\tilde{a}}{\tilde{a}^4E(\tilde{a})\sqrt{1+\bar{R_b}\tilde{a}}}, \label{25}
\end{equation}
where $\bar{R_b}a=3\rho_b/(4\rho_r)$, $\bar{R_b}=31500\omega_b(T_{\mathrm{CMB}}/2.7\mathrm{K})^{-4}$ and we have assumed the CMB temperature $T_{\mathrm{CMB}}=2.7225\mathrm{K}$. It is worth noting that we cannot neglect the effect of radiation when using CMB data to constrain dark energy. Its contribution can be obtained through the so-called matter-radiation equality relation $\Omega_{r0}=\Omega_{m0}/(1+z_{eq})$, where $z_{eq}=2.5\times10^4\Omega_{m0}h^2(T_{\mathrm{CMB}}/2.7\mathrm{K})^{-4}$.   

Subsequently, the decoupling redshift $z_\star$ is calculated by the following fitting formula \cite{39}
\begin{equation}
z_\star = 1048(1+g_1\omega_m^{g_2})(1+0.00124\omega_b^{-0.738}) , \label{26}
\end{equation}
where $\omega_m=\Omega_{m0}h^2$ and 
\begin{equation}
g_1 = \frac{0.0783\omega_b^{-0.238}}{1+39.5\omega_b^{0.763}}, \label{27}
\end{equation}
\begin{equation}
g_2 = \frac{0.560}{1+21.1\omega_b^{1.81}}. \label{28}
\end{equation}
To perform the common $\chi^2$ statistics, we express $\chi^2$ for CMB data as follows 
\begin{equation}
\chi^2 = (\mathbf{v}_{th}-\mathbf{v}_{obs})\mathcal{C}^{-1}(\mathbf{v}_{th}-\mathbf{v}_{obs})^{tr}, \label{29}
\end{equation}
where the subscript $tr$ represents the transpose of a vector or a matrix, $\mathcal{C}$ is the covariance matrix, and $\mathbf{v}_{th}$ and $\mathbf{v}_{obs}$ denote the theoretical and observational values of data vector $\mathbf{v}=(R, \, l_a, \, \omega_b, \, n_s)^{tr}$. Specifically, for a spatially flat universe,  $\mathbf{v}_{obs} = (1.74963,\, 301.80845, \, 0.02237, \, 0.96484)^{tr}$ and  

\begin{equation}
\mathcal{C}=10^{-8}\times\left(
\begin{array}{cccc} 
1598.9554 & 17112.007 & -36.311179 & -1122.4683      \\ 
17112.007 & 811208.45 & -494.79813 & -11925.120           \\ 
-36.311179 & -494.79813 & 2.1242182 & 23.779841             \\
-1122.4683 & 11925.120 & 23.779841 & 1725.4040
\end{array}
     \right). \label{30}
\end{equation}

Determining the mass and species of neutrinos is a very important task in the fields of particle physics and cosmology. Combining BAO data with the latest CMB data, the mass sum of three active neutrinos $\Sigma m_\nu$ and the effective number of relativistic species $N_{eff}$ have been, respectively, tighten to $\Sigma m_\nu < 0.12$ eV and $N_{eff}= 2.99^{+0.34}_{-0.33}$ at the $2\sigma$ confidence level by the Planck collaboration \cite{2}. Since these two neutrino parameters have direct impacts on the sound horizon when the universe is radiation-dominated, they also have effects on $H_0$. Therefore, we also attempt to check whether changes in the neutrino sector can help us alleviate the $H_0$ discrepancy in $f(T)$ gravity. Through the energy density of radiation after electron-positron annihilation \cite{40}, $N_{eff}$ can be defined as    
\begin{equation}
\rho_r=\rho_\gamma\left[ 1+N_{eff}\frac{7}{8}\left(\frac{4}{11}\right)^{\frac{4}{3}}   \right], \label{31}
\end{equation}
where $\rho_\gamma$ denotes the energy density of a photon. If considering the effects of neutrinos on the CMB spectrum, for a flat universe, the authors in Ref.\cite{41} also give the corresponding data vector  $\mathbf{v}_{obs} = (1.7661,\, 301.7293, \, 0.02191, \, 0.1194, \, 2.8979)^{tr}$ and  
\begin{equation}
\mathcal{C}=10^{-8}\times\left(
\begin{array}{ccccc}
33483.54 & -44417.15 & -515.03 &-360.42 &-274151.72        \\
-44417.15 &4245661.67 & 2319.46 & 63326.47 & 4287810.44    \\
-515.03 & 2319.46 & 12.92 & 51.98 & 7273.04                 \\
-360.42 & 63326.47 & 51.98 & 1516.28 & 92013.95            \\
-274151.72 & 4287810.44 & 7273.04 & 92013.95 & 7876074.608    
\end{array}
\right). \label{32}
\end{equation}
Note that the data vector $\mathbf{v}$ here has been changed to $\mathbf{v}=(R, \, l_a, \, \omega_b, \, \omega_c, N_{eff})^{tr}$. 

For the purpose to perform conveniently Bayesian parameter estimation for three $f(T)$ models, we employ the online package {\it EMCEE} \cite{42}, which is an extensible pure-python Affine Invariant Markov chain Monte
Carlo (MCMC) Ensemble sampler. Meanwhile, to analyze the MCMC chains, we take the public package {\it GetDist} \cite{43}.

In order to check the validity of distance prior method, we constrain the $\Lambda$CDM model and see whether the results from Planck collaboration \cite{2} can be recovered. The corresponding marginalized constraints on $\Lambda$CDM are shown in Fig. \ref{f1} and Tab.\ref{t1}. One can easily that the constraining results is very consistent with those given by the Planck Team. Therefore, the above data and method can be used to constrain $f(T)$ theories.

\begin{table}[ph]
	\renewcommand\arraystretch{1.5}
	\caption{The constraining results of free parameters of five different cosmological models from Planck-2018 CMB data. Particularly, we quote $2\sigma$ upper bounds on the parameters $b$, $N_{eff}$ and $\Sigma m_\nu$.}
	{\begin{tabular}{@{}cccccc@{}} \toprule
			Parameters            &$\Lambda$CDM           & M1      &M2           &M3           &M1$\nu$                     \\ \colrule
			$H_0$       &67.35$\pm$0.54  &  66.51$\pm$3.65    & 67.11$\pm$0.56    &  67.12$\pm$0.56   &   66.52$\pm$3.80                  \\  
			$\Omega_{m0}$        &$0.315\pm0.007$  &$0.324\pm0.032$  &$0.318\pm0.007$ &$0.317\pm0.007$ &$0.319\pm0.037$                          \\
			$\Omega_{b0}$                  &$0.0493\pm0.0006$   &$0.0506\pm0.0048$                     &$0.0493^{+0.0006}_{-0.0008}$             &$0.0496^{+0.0008}_{-0.0006}$             &$0.0502^{+0.0059}_{-0.0048}$                        \\
			$b$       &---      &$0.05\pm0.19$                        &$<0.217 \,(2\sigma)$                      &$<0.215 \,(2\sigma)$           &$0.07\pm0.21$                                        \\
			$N_{eff}$           &---                        &---                      &---            &---             &$3.04^{+0.41}_{-0.45}\,(2\sigma)$                          \\
			$\Sigma m_\nu$                  &---                        &---                      &---            &---             &$<0.50 \,(2\sigma)$                        \\
			\botrule
		\end{tabular}
		\label{t1}}
\end{table}

\section{Results}

Our marginalized constraining results of three $f(T)$ models are displayed in Figs.\ref{f2}-\ref{f5} and Tab.\ref{t1}. In light of constraints on the distortion parameter $b$ in three scenarios, we find that there is no any departure from the standard cosmology under the framework of GR, and that the constraining results in this analysis are consistent with those in Refs.\cite{31,32,33,a1,a2,a3}. It is very interesting that current $H_0$ tension can be effectively resolved from $4.4\sigma$ to $1.9\sigma$ in the power-law model M1. However, two exponential models M1 and M2 can hardly alleviate the $H_0$ tension and the constraining results of them is very close to those of $\Lambda$CDM using CMB data. Based on the fact that M1 can effectively mitigate the $H_0$ tension, we attempt to go for a further step to alleviate this tension by considering the effects of free-streaming neutrinos in the universe. As a consequence, for the first time, we place constraints on $\Sigma m_\nu$ and $N_{eff}$ in $f(T)$ gravity. For a degenerate hierarchy as taken by the Planck team, we find that the constraint on $b$ in M1$\nu$ is naturally a little looser than that in $M1$, and that the $2\sigma$ error of effective number of relativistic species $N_{eff}=3.04^{+0.41}_{-0.45}$ and $2\sigma$ upper bound on the mass sum of three active neutrinos $\Sigma m_\nu<0.50$ eV is larger than the prediction $N_{eff}=2.89^{+0.36}_{-0.38}$ and $\Sigma m_\nu<0.24$ eV given by the Planck collaboration \cite{2}, respectively. Specially, the improvement in resolving $H_0$ tension in M1$\nu$ is just small from $1.9\sigma$ to $1.8\sigma$ relative to M1. To show the alleviation of $H_0$ tension in $f(T)$ gravity better, we plot the $H_0$-$b$ contour for M1 and M1$\nu$. From Fig.\ref{6}, it is easy to see that the addition of neutrinos enlarges the parameter space but does not give a obvious enlargement in $H_0$ direction.   

A very important task in $f(T)$ gravity is to study the degeneracy between the distortion parameter $b$ and other cosmological parameters. In Fig.\ref{2}, for M1, one can easily find that $H_0$ is strongly anti-correlated with $b$, which indicated that the universe has a larger expansion rate with decreasing $b$. One the contrary, $b$ is positively correlated with $\Omega_{m0}$ and $\Omega_{b0}$, which implies that matter and baryon densities of the universe increases with increasing $b$.
Very different from M1, in M2 and M3, $b$ is still strongly degenerated with other parameters. This tells us that, in M2 and M3, high redshift information indicates that the parameter $b$ is very insensitive to the cosmic expansion rate $H_0$. 

Note that previous works \cite{31,32,33,34} also obtain the similar results for M2 and M3 by using low redshift data. It is very strange that why M1 can resolve the $H_0$ tension but M2 and M3 cannot. This issue has always been not noticed for a long time. In the following analysis, we shall explain this in a simple way. The most straightforward to address this issue is to study the effect of variation of $H_0$ on the distortion factor $y(z,\,b,\,\Omega_{m0},\Omega_{r0},\,H_0)$. Firstly, we choose $H_0=70$ km s$^{-1}$ Mpc$^{-1}$ as the baseline value and assume $b=0.1$, $\Omega_{m0}=0.3$ and $\Omega_{r0}=8.47\times10^{-5}$ for three $f(T)$ models, and then define the relative difference of distortion factor $\epsilon$ as 
\begin{equation}
\epsilon\equiv\frac{\Delta y}{y}=\frac{y(z,0.1,0.3, 8.47\times10^{-5},H_0)-y(z,0.1,0.3,8.47\times10^{-5},70)}{y(z,0.1,0.3,8.47\times10^{-5},70)}=\frac{y(z,0.1,0.3,8.47\times10^{-5},H_0)}{y(z,0.1,0.3,8.47\times10^{-5},70)}-1.          \label{33}
\end{equation} 

The numerical results are displayed in Fig.\ref{f7}. One can easily find that the $\epsilon$ value always keeps zero for M2 and M3, while it increases gradually with increasing redshift for M1. It indicates that the distortion factor $y$ is insensitive to the $H_0$ variation at all redshifts for M2 and M3, but becomes more and more sensitive to the $H_0$ value with increasing redshift for M1. This is the reason why the power-law model M1 can resolve the $H_0$ tension more efficiently than exponential models M2 and M3 do. Actually, the insensitivity of $H_0$ to $y$ for M2 and M3 can also be seen from Eq.(\ref{19}) and Eq.(\ref{22}). When $z$ approaches $z_\star\sim1090$, for given parameters $b$, $\Omega_{m0}$ and $\Omega_{r0}$, the dimensionless Hubble parameter $E(z)$ tends to be very large, which naturally leads to $y\approx1$. Differently, for M1, $y$ and $E(z)$ monotonically increase with increasing $z$. Furthermore, by comparing Eq.(\ref{16}) with Eq.(\ref{19}) and Eq.(\ref{22}), we obtain a conclusion that whether a viable $f(T)$ theory can mitigate the $H_0$ tension depends on the mathematical structure of $y$, i.e., the specific choice of distortion factor.  

As a complementary analysis, we also investigate the evolutionary behaviors of effective EoS of dark energy in three $f(T)$ models in Fig.\ref{f8}. For M1, we find that when adopting a larger redshift $z$, the EoS of dark energy tends to depend linearly on the distortion paarameter $b$, and that when adopting a more positive or negative value of $b$, the EoS not only monotonically increases but also deviates from -1 more largely. Using the same analysis method, for M2 and M3, we find that when taking a larger value of $z$, their EoSs tend to have the same behavior as EoS of $\Lambda$CDM with increasing $b$, and that when fixing $b$, their EoSs will converge to -1 quickly, regardless of values of $b$. This indicates that M2 and M3 have the same behaviors as $\Lambda$CDM at high redshifts, which can also help explain why M2 and M3 cannot relieve the $H_0$ tension at all.             

It is worth noting that the alleviation of $H_0$ in M1 is based on the fact that we have obtained a lower mean value of $H_0$ but with a larger uncertainty than those in $\Lambda$CDM by using the Planck CMB distance information. This implies that the free parameter $b$ in M1 is insensitive to CMB distance data, enlarge the parameter space and consequently leads to a large growth of uncertainty of $H_0$. To be more specific, the insensitiveness could be ascribed to the power law form $(-T)^b$, where $b$ is the power and, generally, could not be well constrained by CMB data. We think that it is still hard to compress the error of $H_0$ in M1, even if future CMB data has a higher precision than Planck. In order to obtain a higher mean value and lower error of $H_0$ than those in $\Lambda$CDM, one may consider some useful power law forms of torsional scalar $T$ or other specific $f(T)$ functions. As described above, our results provide a good clue for theoreticians to construct a physically reasonable $f(T)$ function, which can be well constrained by observations and give a great alleviation of the Hubble tension.       

\section{Conclusions}
Motivated by the large discrepancy in measurements of $H_0$ between local and global probes, we investigate whether the teleparallel gravity equivalent to GR could be a better solution to describe the present days observations or at least could alleviate the $H_0$ tension. Specifically, in this work we study and place constraints on three popular $f(T)$ models in light of the Planck-2018 CMB data release.   

We find that the $f(T)$ power-law model can alleviate the $H_0$ tension from $4.4\sigma$ to $1.9\sigma$ level, while the $f(T)$ model of two exponential fail to resolve this inconsistency. 

For the first time, using the Planck-2018 temperature, polarization and lensing data, we obtain constraints on the effective number of relativistic species $N_{eff}$ and on the sum of the masses of three active neutrinos $\Sigma m_\nu$ in $f(T)$ gravity. We find that the constraints obtained are looser than those given by the Planck collaboration under the assumption of $\Lambda$CDM. The introduction of massive neutrinos into the cosmological model does not improve the $H_0$ tension in the case of the exponential-law model. However, for the $f(T)$ power-law model, it does indeed alleviate  the $H_0$ tension. Very interestingly, we find that whether a viable $f(T)$ theory can mitigate the $H_0$ tension depends on the mathematical structure of the distortion factor $y(z,\,b)$. These results could provide a clue for theoreticians to write a physically motivated expression of $f(T)$ function.       

\section{Acknowledgements}
DW thanks Xiaodong Li and Ji Yao for useful communications in HOUYI workshop. DW also thanks Shihong Liao and Jiajun Zhang for helpful discussions on dark matter. DW is supported by the Super Postdoc Project of Shanghai City. DFM thanks the Research Council of Norway for their support and the UNINETT Sigma2 -- the National Infrastructure for High Performance Computing and  Data Storage in Norway.


\begin{thebibliography}{99}

\bibitem{1}
M. Tanabashi {\it et al.} [Particle Data Group],
Phys.\ Rev.\ D {\bf 98}, 030001 (2018).

\bibitem{2}
N.~Aghanim {\it et al.} [Planck Collaboration],
arXiv:1807.06209 [astro-ph.CO].

\bibitem{3}
C. L. Bennett {\it et al.} [WMAP Collaboration], Astrophys. J. Suppl. Ser. {\bf 208}, 20 (2013).

\bibitem{4}
P. Ade {\it et al.} [Planck Collaboration], Astron. Astrophys. {\bf 571}, A16 (2014).

\bibitem{5}
A. G. Riess et al. [Supernova Search Team], Astron. J. {\bf 116}, 1009 (1998).

\bibitem{6}
S. Perlmutter, et al. [Supernova Cosmology Project], Phys. Rev. Lett. {\bf 83}, 670 (1999).

\bibitem{7}
C.~Blake and K.~Glazebrook,
Astrophys.\ J.\  {\bf 594}, 665 (2003).

\bibitem{8}
H.~J.~Seo and D.~J.~Eisenstein,
Astrophys.\ J.\  {\bf 598}, 720 (2003).

\bibitem{9} 
H.~Hildebrandt {\it et al.},
Mon.\ Not.\ Roy.\ Astron.\ Soc.\  {\bf 465}, 1454 (2017).

\bibitem{10} 
T.~M.~C.~Abbott {\it et al.} [DES Collaboration],
Phys.\ Rev.\ D {\bf 98}, 043526 (2018).

\bibitem{11} 
T.~Hamana {\it et al.} [HSC Collaboration],
Publ.\ Astron.\ Soc.\ Jap.\  {\bf 72}, no. 1, (2020). 

\bibitem{12} 
S.~Weinberg,
Rev.\ Mod.\ Phys.\  {\bf 61}, 1 (1989).

\bibitem{13} 
A.~G.~Riess, S.~Casertano, W.~Yuan, L.~M.~Macri and D.~Scolnic,
Astrophys.\ J.\  {\bf 876}, 85 (2019).

\bibitem{14} 
N.~Aghanim {\it et al.} [Planck Collaboration],
Astron.\ Astrophys.\  {\bf 596}, A107 (2016).

\bibitem{15} 
R.~A.~Battye, T.~Charnock and A.~Moss,
Phys.\ Rev.\ D {\bf 91}, 103508 (2015).

\bibitem{16} 
E.~Macaulay, I.~K.~Wehus and H.~K.~Eriksen,
Phys.\ Rev.\ Lett.\  {\bf 111}, 161301 (2013).

\bibitem{17} 
J.~L.~Bernal, L.~Verde and A.~G.~Riess,
JCAP {\bf 1610}, 019 (2016).

\bibitem{18} 
G.~Benevento, W.~Hu and M.~Raveri,
arXiv:2002.11707 [astro-ph.CO].

\bibitem{19} 
D.~Wang and X.~H.~Meng,
arXiv:1709.04141 [astro-ph.CO].

\bibitem{20} 
S.~Kumar, R.~C.~Nunes and S.~K.~Yadav,
Eur.\ Phys.\ J.\ C {\bf 79}, 576 (2019).

\bibitem{21} 
V.~Poulin, T.~L.~Smith, T.~Karwal and M.~Kamionkowski,
Phys.\ Rev.\ Lett.\  {\bf 122}, 221301 (2019).

\bibitem{22} 
D.~Wang and X.~H.~Meng,
Phys.\ Rev.\ D {\bf 96}, 103516 (2017).

\bibitem{23} 
D.~Wang, Y.~J.~Yan and X.~H.~Meng,
Eur.\ Phys.\ J.\ C {\bf 77}, 660 (2017).

\bibitem{24} 
J.~C.~Hill, E.~McDonough, M.~W.~Toomey and S.~Alexander,
arXiv:2003.07355 [astro-ph.CO].

\bibitem{25} 
S.~Ghosh, R.~Khatri and T.~S.~Roy,
arXiv:1908.09843 [hep-ph].

\bibitem{26} 
A.~De Felice, C.~Q.~Geng, M.~C.~Pookkillath and L.~Yin,
arXiv:2002.06782 [astro-ph.CO].

\bibitem{27} 
W.~E.~V.~Barker, A.~N.~Lasenby, M.~P.~Hobson and W.~J.~Handley,
arXiv:2003.02690 [gr-qc].

\bibitem{28} 
R. Aldrovandi and J. G. Pereira, {\it Teleparallel Gravity: An Introduction}, Springer, Dordrecht
(2013).

\bibitem{29} 
Y.~F.~Cai, S.~Capozziello, M.~De Laurentis and E.~N.~Saridakis,
Rept.\ Prog.\ Phys.\  {\bf 79}, 106901 (2016).

\bibitem{30} 
G.~R.~Bengochea and R.~Ferraro,
Phys.\ Rev.\ D {\bf 79}, 124019 (2009).

\bibitem{31} 
P.~Wu and H.~W.~Yu,
Phys.\ Lett.\ B {\bf 693}, 415 (2010).

\bibitem{32} 
S.~Nesseris, S.~Basilakos, E.~N.~Saridakis and L.~Perivolaropoulos,
Phys.\ Rev.\ D {\bf 88}, 103010 (2013).

\bibitem{33} 
R.~C.~Nunes, S.~Pan and E.~N.~Saridakis,
JCAP {\bf 1608}, 011 (2016).

\bibitem{34} 
F.~K.~Anagnostopoulos, S.~Basilakos and E.~N.~Saridakis,
Phys.\ Rev.\ D {\bf 100}, 083517 (2019).

\bibitem{a1} 
  A.~Awad, W.~El Hanafy, G.~G.~L.~Nashed and E.~N.~Saridakis,
  JCAP {\bf 1802}, 052 (2018).
  
  
\bibitem{a2} 
  V.~F.~Cardone, N.~Radicella and S.~Camera,
  Phys.\ Rev.\ D {\bf 85}, 124007 (2012).

\bibitem{a3} 
  S.~Camera, V.~F.~Cardone and N.~Radicella,
  Phys.\ Rev.\ D {\bf 89}, 083520 (2014).



\bibitem{35}
R. Weitzenb\"{o}ck, {\it Invariantentheorie} (Gronningen:
Noordhoff) (1923).

\bibitem{36} 
E.~V.~Linder,
Phys.\ Rev.\ D {\bf 81}, 127301 (2010),
Erratum: [Phys.\ Rev.\ D {\bf 82}, 109902 (2010)].

\bibitem{37} 
K.~Bamba, C.~Q.~Geng, C.~C.~Lee and L.~W.~Luo,
JCAP {\bf 1101}, 021 (2011).

\bibitem{b1}
B.~Li, T.~P.~Sotiriou and J.~D.~Barrow,
Phys. Rev. D \textbf{83}, 104017 (2011).

\bibitem{b2}
C.~Li, Y.~Cai, Y.~F.~Cai and E.~N.~Saridakis,
JCAP \textbf{10}, 001 (2018).

\bibitem{b3}
A.~Golovnev and T.~Koivisto,
JCAP \textbf{11}, 012 (2018).






\bibitem{38} 
Z.~Zhai and Y.~Wang,
JCAP {\bf 1907}, 005 (2019).

\bibitem{39} 
W.~Hu and N.~Sugiyama,
Astrophys.\ J.\  {\bf 471}, 542 (1996).

\bibitem{40} 
J.~Lesgourgues and S.~Pastor,
New J.\ Phys.\  {\bf 16}, 065002 (2014).

\bibitem{41} 
Z.~Zhai, C.~G.~Park, Y.~Wang and B.~Ratra,
arXiv:1912.04921 [astro-ph.CO].

\bibitem{42} 
D.~Foreman-Mackey, D.~W.~Hogg, D.~Lang and J.~Goodman,
Publ.\ Astron.\ Soc.\ Pac.\  {\bf 125}, 306 (2013).

\bibitem{43} 
A.~Lewis,
arXiv:1910.13970 [astro-ph.IM].


































\end{thebibliography}
\end{document}